\documentclass[prl,twocolumn,amssymb,superscriptaddress]{revtex4-2}
\usepackage{hyperref,amsmath,graphicx,subfigure}
\usepackage{geometry}
\newcommand{\arXiv}[1]{\href{http://www.arXiv.org/abs/#1}{arXiv:#1}}
\renewcommand{\doi}[2]{\href{http://dx.doi.org/#2}{#1}}
\newcommand{\beq}{\begin{equation}}
\newcommand{\eeq}{\end{equation}}

\DeclareMathOperator{\Tr}{Tr}

\begin{document}

\title{Multiseed Krylov Complexity}

\author{Ben Craps} 
\affiliation{Theoretische Natuurkunde, Vrije Universiteit Brussel (VUB) and International Solvay Institutes, Brussels, Belgium}
\author{Oleg Evnin}
\affiliation{High Energy Physics Research Unit, Faculty of Science, Chulalongkorn University, Bangkok, Thailand}
\affiliation{Theoretische Natuurkunde, Vrije Universiteit Brussel (VUB) and International Solvay Institutes, Brussels, Belgium}
\author{Gabriele Pascuzzi}
\affiliation{Theoretische Natuurkunde, Vrije Universiteit Brussel (VUB) and International Solvay Institutes, Brussels, Belgium}

\begin{abstract}
Krylov complexity is an attractive measure for the rate at which quantum operators spread in the space of all possible operators under dynamical evolution.
One expects that its late-time plateau would distinguish between integrable and chaotic dynamics, but its ability to do so depends precariously
on the choice of the initial seed. We propose to apply such considerations not to a single operator, but simultaneously
to a collection of initial seeds in the manner of the block Lanczos algorithm. We furthermore suggest that this collection should comprise all simple (few-body) operators in the theory, which echoes the applications of Nielsen complexity to dynamical evolution. The resulting construction,
unlike the conventional Krylov complexity, reliably distinguishes integrable and chaotic Hamiltonians without any need for fine-tuning.
\end{abstract}

\maketitle

Ever since its introduction in \cite{K}, Krylov complexity has been one of the key approaches to manifesting the information-theoretic content of quantum dynamics.
The idea is to track how rapidly the Heisenberg evolution of a given initial quantum operator explores different directions in the space of all operators.
If many extra directions enter the game rapidly, the evolution cannot be approximated well by an effectively truncated subspace, and is, in this sense, {\it complex}.
A recent comprehensive review may be found in \cite{review}. Similar ideas have been explored for the Schr\"odinger evolution of states, instead of the Heisenberg evolution of operators, starting with \cite{Kstate}.

Attractive as it is, the practical performance of Krylov complexity has met some challenges, and our goal here is to present an upgrade that addresses these challen\-ges. A key property one expects from complexity measures of quantum evolution is that they should assign smaller values to integrable/solvable systems than to generic/chaotic systems. If successful, this would give a mathematical expression to the intuitive notion that solved problems are easier than unsolved ones. Krylov complexity tends to grow at early times and saturate at a plateau at late times \cite{satur1,satur2}, and the height of this late-time plateau is one possible indicator of how complex a system is. This program has been seen to work well in some cases, and some general principles have been spelled out for how integrability may
reduce the height of the late-time plateau \cite{Kint}. However, the success of this approach depends on the choice of the initial operator, for which there has been no systematic understanding up to this point. For example, in \cite{Kdep}, the performance of the late-time plateau as an indicator of integrability of a spin chain is broken by choosing a particular spatial projection of a single-site spin as the initial seed, while another spatial projection of the same single-site spin leads to the desired performance. Further discussions of the late-time plateau and its dependence on the initial operator can be found in \cite{plateau, plateau1, plateau2, hump}.

Setting aside for a moment the undesirable sensitivity of Krylov complexity performance to the choice of the initial seed, there is a broader conceptual problem:
the need to choose any initial seed at all is hardly appealing. In the end, one would like to obtain a characterization of a physical system in terms of whether its evolution is simple or complex. Standard Krylov complexity, however, takes as its input a Hamiltonian and an initial seed. The result depends crucially on the seed. For example,
taking a conserved operator as the seed results in vanishing Krylov complexity for all systems. Is there a way to condense all these operator-by-operator evaluations into a statement about the system that does not depend on the initial seed? It has often been implicit in the literature that the initial seed should be a `simple' operator of sorts, as in the single-spin example mentioned in the previous paragraph, but this is seen as a practical choice within each concrete setup, without being inherent to the underlying definitions. We will incorporate the notion of simple operators systematically.

Choosing an appropriate set of simple operators has been at the heart of applications of Nielsen complexity to quantum evolution \cite{evolcompl1,evolcompl2,bound,complint}, an approach developed in parallel with Krylov complexity, starting from a rather different set of first principles. (Relations between Krylov and Nielsen complexity have been explored in \cite{dilaton,Lvetal,notdistance,KrNl}.) Nielsen complexity takes as its input a quantum Hamiltonian and a collection of Hermitian operators designated as `simple' (they define `easy' directions in the space of unitaries where the evolution unfolds). The notion of what is simple is an essential input also in the context of standard computational complexity theory, which asks how many elementary/simple/fast operations are needed to execute the desired algorithm, and this notion has migrated from there to the definition of Nielsen complexity. The choice of simple operators for a quantum system is made from inspecting its degrees of freedom, with a prominent role played by few-body operators, that is, those that only act on a few degrees of freedom at once. For example, for a spin chain, one may choose to label as simple all those operators that act on a single spin, or those that only act on two adjacent spins, etc. In this way, a characterization of the system is produced that does not refer to the evolution of a single chosen operator in the way Krylov complexity does, since specifying the set of all simple operators is guided by clear physical principles.

We will adopt a similar framework in our upgrade of Krylov complexity. Instead of applying the protocol of \cite{K} to a single initial operator, we will modify it by including multiple initial operators, chosen according to the same principles as in the work on Nielsen complexity, and observe how this set spreads out dynamically to include more complex operators. While the formulation of \cite{K} relies on the Lanczos algorithm for matrix tridiagonalization,
our {\it multiseed} upgrade is naturally powered by the block Lanczos algorithm previously discussed in the mathematical literature on numerical methods
\cite{blockL}. We will see that, paired with a natural specification of the block Lanczos seed as all few-body operators in the theory, this setup reliably distinguishes integrable and chaotic evolution without any further fine-tuning.

{\it Lanczos algorithm with single and multiple seeds.---} We start with a brief review of the standard single-seed Krylov complexity formulated in terms of the ordinary Lanczos algorithm \cite{hist}, before proceeding with the multiseed complexity and block Lanczos algorithm. To optimize the notation, we represent every operator $\mathcal{O}$ as a `state' in the space of operators, and write it as $|\mathcal{O}\rangle$. The Krylov basis is constructed given an initial {\it seed} operator $|\mathcal{O}_0\rangle$ and the Liouvillian $\mathcal{L} = [{H},\cdot]$. It can be defined as an orthogonalization of the Krylov sequence $\big\{|\mathcal{O}_0\rangle, \mathcal{L}|\mathcal{O}_0\rangle, \mathcal{L}^2|\mathcal{O}_0\rangle, ...\big\}$ with respect to the inner pro\-duct \cite{Frob} $\langle\mathcal{A}|\mathcal{B}\rangle \equiv\Tr [{\mathcal{A}}^{\dagger}{\mathcal{B}}]$. As a consequence of its hermiti\-ci\-ty, the Liouvillian is tridiagonal in the Krylov basis $|\mathcal{O}_j\rangle$:
\begin{align}
\mathcal{L}|\mathcal{O}_j\rangle = b_{j+1}|\mathcal{O}_{j+1}\rangle + a_j |\mathcal{O}_{j}\rangle + b_j|\mathcal{O}_{j-1}\rangle.
\label{Laczstep}
\end{align}
Additionally, for Hermitian seeds, all $a_j$ vanish. Solving for $|\mathcal{O}_{j+1}\rangle$, one arrives at the Lanczos algorithm for constructing the basis, which provides a significant simplification compared to the usual Gram-Schmidt procedure.
For finite-dimensional spaces, for some $j=K-1$, acting on $\mathcal{O}_{K-1}$ with $\mathcal{L}$ will result in an operator that is a linear combination of the previous ones, so that $\mathcal{O}_{K}=0$ and the algorithm terminates.
The resulting Krylov basis of dimension $K$ is, by construction, able to cover the full time evolution of the initial operator $\mathcal{O}_0$:
\beq
    |\mathcal{O}_0(t)\rangle = e^{i\mathcal{L}t}|\mathcal{O}_0\rangle = \sum_{j=0}^{K-1} \phi_j (t) |\mathcal{O}_j \rangle.\vspace{-1mm}
\label{phidecomp}
\eeq
When starting with a local operator, more applications of $\mathcal{L}$ will typically create more nonlocal operators, so that the basis is ordered according to increasing complexity. This motivates defining Krylov complexity as the average `position' of an operator in the Krylov basis, so that having more support on more complex operators is expressed mathematically as higher values of complexity:\vspace{-4mm}
\beq\label{CKdef}
 C_K (t) = \sum_{j=0}^{K-1} j |\phi_j (t) |^2.
\eeq

In applications, Krylov complexity (\ref{CKdef}) shows early-time growth, followed by saturation at a plateau at late times \cite{satur1,satur2}. The height of this plateau can be computed \cite{Kint} as the all-time average of (\ref{CKdef}), yielding
\beq\label{CKav}
\overline{C_{K}}=\sum_{\alpha=0}^{K-1}|\langle \mathcal{O}_0|\omega_\alpha\rangle|^2\sum_{j=0}^{K-1} j  |\langle \mathcal{O}_j|\omega_\alpha\rangle|^2,
\eeq
where $|\omega_\alpha\rangle$ are the eigenstates of the Liouvillian $\mathcal{L}$ restricted to the Krylov space.
(The usage of the all-time average lets one bypass the moment-by-moment dynamical evolution and obtain the late-time saturation value directly.)
The complexity plateau height has been shown to decrease in the presence of integrability for some systems and for some specific initial operators \cite{Kint}. However, this link is not general, and it easily breaks down even for relatively simple systems and local operators \cite{Kdep}. We would like to cure this dependence of the Krylov complexity performance on the choice of initial operator.

Inspired by the successes of Nielsen complexity, as mentioned in the introduction, we attempt to stabilize the performance of the Krylov complexity plateau
by filling the lowest-weight subspace with all the simplest local operators of the system under consideration (e.g.\ 1-body operators), and then using the Liouvillian as before to get progressively more complex subspaces. Specifically, we start with a collection of $m$ seed operators, $\big\{ |\mathcal{O}_{0,0}\rangle,|\mathcal{O}_{0,1}\rangle, ...,|\mathcal{O}_{0,m-1}\rangle \big\}$, which we can normalize and make mutually orthogonal, and which we refer to collectively as $\displaystyle {\Omega}_0$. The Krylov basis will now consist of the orthogonalization of all the operators in the set $\displaystyle \big\{{\Omega}_0, {\mathcal{L}}{\Omega}_0, {\mathcal{L}^2}{\Omega}_0, ...\big\}$, obtained by repeatedly applying the Liouvillian to the initial set. As before, in finite-dimensional spaces, applying $\displaystyle {\mathcal{L}}$ will eventually yield operators that are all linear combinations of the previous ones, which means the construction of the block Krylov basis $\displaystyle \big\{ {\Omega}_0,{\Omega}_1, ...,{\Omega}_{M-1} \big\}$ is complete. The operators in $\displaystyle {\Omega}_J$ are obtained by orthogonalizing those in $\displaystyle {\mathcal{L}^J}{\Omega}_0$ against all previous ones and against each other, and normalizing. An operator can be discarded if it is a linear combination of the previous ones, so that $\displaystyle {\Omega}_J$ contains $p_J$ operators, with $p_J$ non-increasing with $J$:
\beq
{\Omega}_J = \big\{ |\mathcal{O}_{J,0}\rangle,|\mathcal{O}_{J,1}\rangle ,... , |\mathcal{O}_{J, p_J-1}\rangle \big\}.
\eeq
Once again, the construction is simplified due to the hermiticity of the Liouvillian, in a manner analogous to \eqref{Laczstep}, resulting in what has been known as the {\it block Lanczos algorithm} in the mathematical literature on numerical methods \cite{blockL}, but has not been applied to quantum complexity topics up to this point
\cite{blockhop}.

The block Krylov basis, by construction, can cover the time evolution of each of the seed operators (or any linear combination thereof):
\beq
    |\mathcal{O}_{0,n} (t)\rangle = e^{i\mathcal{L}t} |\mathcal{O}_{0,n}\rangle = \sum_{J=0}^{M-1} \sum_{k=0}^{p_J-1} \phi_{J,k}^{(n)} (t) |\mathcal{O}_{J,k} \rangle.
\label{tkryl}
\eeq
Following the intuition that more applications of the Liouvillian result in more complex operators, we assign complexity $J$ in the range $0,\ldots,M-1$ to all operators in $\displaystyle {\Omega}_J$, resulting in the following definition \cite{Krylovbasis} of complexity for the evolution of each individual seed operator $\mathcal{O}_{0,n}$ in $\Omega_0$ indexed by $n$:
\beq
    C_K^{(n)} (t) = \sum_{J=0}^{M-1} \sum_{k=0}^{p_J-1} J | \phi_{J,k}^{(n)} (t) |^2.
\label{multicomp}
\eeq
As for the standard Krylov complexity, one can think of (\ref{multicomp}) as an `average position' of the evolving operator, but now relative to the block-Lanczos iterative level $J$, rather than to the ordinary Krylov basis numbering. With a seed containing $m$ simple local operators in $\displaystyle \Omega_0$, as described above, we define {\it multiseed Krylov complexity} as the average of (\ref{multicomp}) over all these seed operators:
\beq
    C_\text{mult} (t) = \frac{1}{m}\sum_{n=0}^{m-1} C_K^{(n)} (t).
\label{multicompavg}
\eeq
With the weight assignment in (\ref{multicomp}), the complexity only depends on the subspace $\displaystyle\Omega_0$ spanned by the initial seeds, not on the specific basis chosen within $\displaystyle \Omega_0$.
Our main interest is in the late-time plateau of this quantity, captured by the all-time average
\beq
\overline{C_\text{mult}} = \lim_{T \to \infty}\frac{1}{T}\int_{0}^T C_\text{mult} (t) dt.
\label{multicomplate}
\eeq
In analogy to (\ref{CKav}), as developed in \cite{Kint}, this expression is simplified by introducing an orthonormal eigenbasis of $\mathcal{L}$, denoted $|\omega_\alpha\rangle$ with the corresponding eigenvalues $\omega_\alpha$:
\begin{align}
\overline{C_\text{mult}} &= \frac{1}{m}\sum_{n=0}^{m-1} \sum_{J=0}^{M-1} \sum_{k=0}^{p_J-1}\!\!\sum_{\substack{\alpha,\beta \\ \omega_\alpha = \omega_\beta}}\!\! J \langle \mathcal{O}_{0,n} | \omega_\alpha \rangle \langle \omega_\beta | \mathcal{O}_{0,n} \rangle \nonumber\\
    &\hspace{1.5cm}\times \langle \mathcal{O}_{J,k} | \omega_\beta \rangle \langle\omega_\alpha | \mathcal{O}_{J,k} \rangle.
\label{multiplateau}
\end{align}
Our main goal is to demonstrate, for a few standard test systems, that the late-time plateau of the multiseed Krylov complexity defined by (\ref{multiplateau})
performs reliably as an indicator of integrability vs.\ chaos, unlike the late-time plateau of the standard Krylov complexity defined by (\ref{CKav}).\vspace{2mm}

{\it Numerical implementation.---} For any Lanczos-type algorithm --- and these issues are more pronounced for block Lanczos algorithms --- one needs to manage the numerical instabilities, which otherwise quickly build up at each step due to large subtractions in the course of orthogonalization. In general, this is done by explicitly reorthogonalizing the operators after some number of iterative steps. One can choose to reorthogonalize at every step (full reorthogonalization), or find a way to determine when enough error has accumulated to make reorthogonalization necessary. For our purposes, we have found that even for modest operator space dimensions of a few thousand, guaranteeing orthogonality up to the standard machine precision $\sim\!10^{-16}$ using full reorthogonalization is not enough to get an accurate Krylov basis (the output basis is visibly unstable with respect to increasing the arithmetic precision). We therefore use high-precision arithmetic in order to guarantee orthogonality with as much precision as needed. Furthermore, given the high numerical cost of full reorthogonalization, we opted for the partial reorthogonalization of \cite{blocklanPRO}: after every Lanczos step, one can estimate the current level of orthogonality of the basis, and only reorthogonalize if some threshold has been reached \cite{notePRO}. Specifically, the operators generated in the current and previous steps are both reorthogonalized against all previous ones and among themselves. Finally, even for the steps when reorthogonalization is not needed, the current list is orthogonalized against the previous two. If the precision of operations is $\epsilon$, the threshold for reorthogonalization is set to $\sqrt{\epsilon}$, so orthogonality of the final basis is guaranteed only up to $\sqrt{\epsilon}$. Though this approach uses more memory, it is more time-efficient than using full reorthogonalization at precision $\sqrt{\epsilon}$ directly, since it only reorthogonalizes when needed. The rest of the computations involved in calculating \eqref{multiplateau} do not suffer from instabilities, and were therefore performed at standard precision. We have made all of our numerical scripts public \cite{repos}.

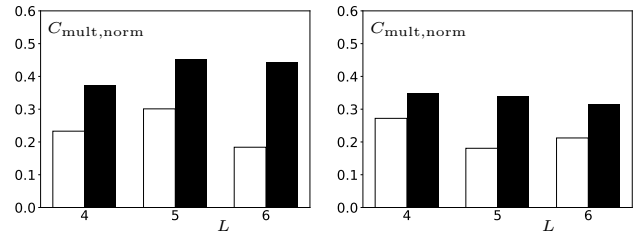
\begin{figure}[t]
\centering
\mbox{\subfigure{\includegraphics[width=0.23\textwidth]{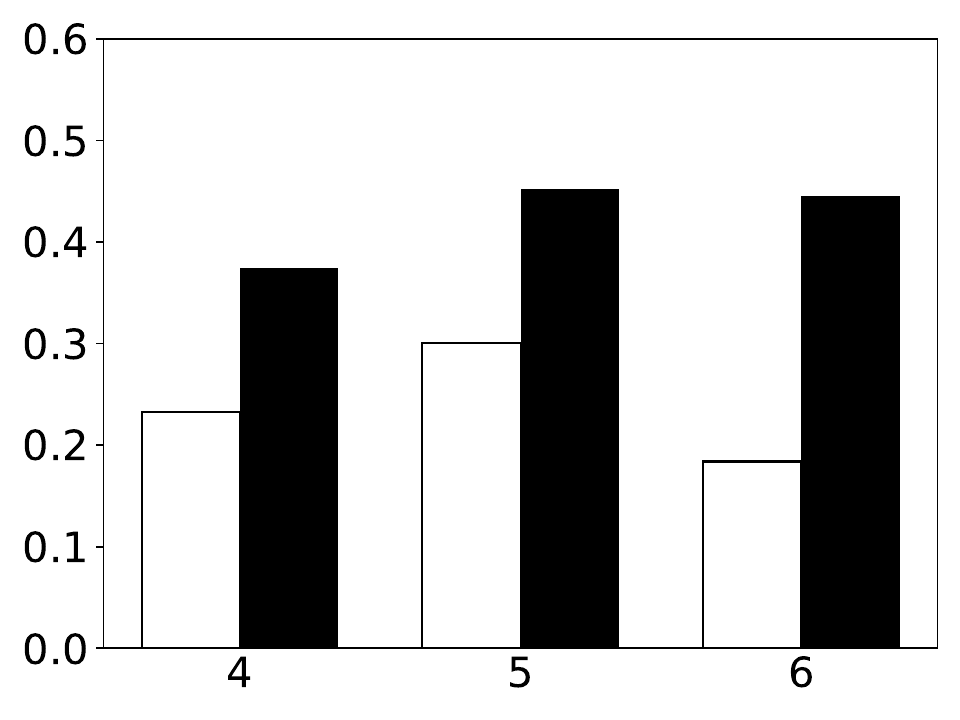}}\,
 \subfigure{\includegraphics[width=0.23\textwidth]{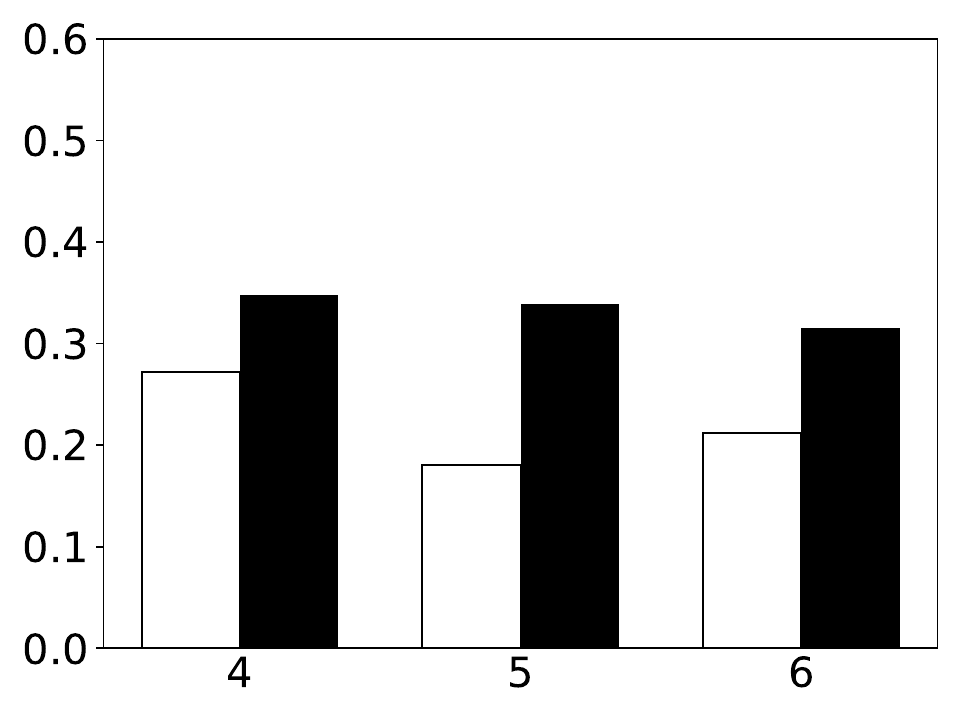}}}
\begin{picture}(0,0)
\put(-40,0){$\scriptstyle L$}
\put(-163,0){$\scriptstyle L$}
\put(-103,75){$\scriptstyle C_\text{mult,norm}$}
\put(-224,75){$\scriptstyle C_\text{mult,norm}$}
\end{picture}
\caption{The multiseed complexity plateau {\bf (left)} for the Ising Hamiltonian \eqref{hising}, with $(h_x,h_z)$ taken as $(-1.05,0)$ and $(-1.05,0.5)$ for the integrable and chaotic versions, respectively, {\bf (right)} for the XYZ Hamiltonian \eqref{hxyz} with $(J_x,J_y,J_z)=(-0.35,0.5,-1)$ and the magnetic field set to $h_z = 0$ for the integrable chain and $h_z = 0.8$ for the chaotic one. Chains with $L=4$, 5, 6 sites are considered. The integrable values are depicted as white bars and the chaotic ones as black bars, seen to be always higher than their white counterparts.
To plot this and all subsequent results, we normalize \eqref{multiplateau} for both integrable and chaotic versions of the system at each given size by the maximal iterative level $M-1$ reached by the block Lanczos process, $C_\text{mult,norm}={C_\text{mult}}/[\max(M_\text{integrable},M_\text{chaotic})-1]$, so that the resulting quantity takes values between 0 and 1. \vspace{-5mm}}
\label{fig:im}
\end{figure}
{\it Multiseed complexity of spin chains.---} We now turn to examining the performance of the multiseed Krylov complexity plateau (\ref{multiplateau}) in a variety of concrete models, starting with spin chains that provide an exemplary laboratory for quantum chaos studies. The spin chains we consider are the mixed-field Ising chain and the spin $1/2$ XYZ Heisenberg chain. We use periodic boundary conditions for both chains, and identify site $L+1$ with the first site. The Hamiltonian describing the Ising model is
\beq
    H_{\text{Ising}} = -\sum_{j=1}^{L} [S_z^{(j)}S_z^{(j+1)} + h_x S_x^{(j)} + h_z S_z^{(j)}].
\label{hising}
\eeq
The system is integrable on both the $h_x=0$ and the $h_z=0$ lines \cite{JW, freefermions}. The first case is trivial, the Hamiltonian being built from commuting terms, so we use the nontrivial case $h_z=0$ as our representative integrable Hamiltonian. Around $(h_x,h_z)=(-1.05,0.5)$, the model exhibits strongly chaotic behavior \cite{stronglychaotic,Lyapunov}, which we choose as our representative  chaotic Hamiltonian. 
For the XYZ chain, we take the usual Hamiltonian along with a magnetic field in the z-direction:
\begin{align}
    H_{\text{XYZ}} = &\sum_{j=1}^{L} [J_x S_x^{(j)}S_x^{(j+1)}\! + \!J_y S_y^{(j)}S_y^{(j+1)}\! + \!J_z S_z^{(j)}S_z^{(j+1)}\nonumber \\
    &-h_z S_z^{(j)}].
\label{hxyz}
\end{align}
For $h_z=0$, the above Hamiltonian is integrable for any values of $J_x$, $J_y$, $J_z$. We take these to be all different for the integrable representative of this model, $(J_x,J_y,J_z)=(-0.35,0.5,-0.1)$. For the chaotic counterpart we set $h_z=0.8$, which is enough to be firmly in the chaotic regime \cite{complint}. 

In all the above cases, we pick the `simple' operators forming the initial seed ${\Omega}_0$ to be the collection of all single-site spin operators. The performance of our algorithm is summarized in Fig.~\ref{fig:im}, showing that it consistently assigns smaller complexity to the integrable cases.

{\it Multiseed complexity of quantum resonant systems.---} In addition to spin chains, we consider quantum resonant systems, which are a class of bosonic models
with quartic interactions typical of many-body physics:
\beq\label{QRSdef}
    H_{\text{QRS}} = \frac{1}{2} \sum_{\substack{n,m,k,l=0 \\ n+m=k+l}}^{\infty} C_{nmkl} a_n^\dagger a_m^\dagger a_k a_l,
\eeq
$C_{nmkl} = C_{klnm} = C_{nmlk}$, $[a_n, a_m^\dagger] = \delta_{nm}$. These systems have been introduced systematically in \cite{quantres}, while earlier applications to trapped interacting bosons, with a specific choice of $C_{nmkl}$, can be found in \cite{qLLL}. An advantage of these systems is that they are not only very tractable numerically, but also possess semiclassical limits in the form of field-theoretic Hamiltonians with rich and diverse dynamics \cite{resrev},
thus providing fertile grounds for quantum chaos studies. Essential for the simplicity of these models is the presence of two conservation laws:\vspace{-1mm}
\beq
    N = \sum_{n=0}^{\infty} a_n^\dagger a_n, \quad M = \sum_{n=1}^{\infty} n\, a_n^\dagger a_n.
\eeq
The Hamiltonian has vanishing matrix elements between states with distinct values of $(N,M)$, but one can easily check that each such $(N,M)$-block is spanned by a finite number of Fock vectors. Diagonalizing (\ref{QRSdef}) is therefore reduced to diagonalizing finite-sized matrices.

\begin{figure}[t]
\centering
\mbox{\subfigure{\includegraphics[width=0.23\textwidth]{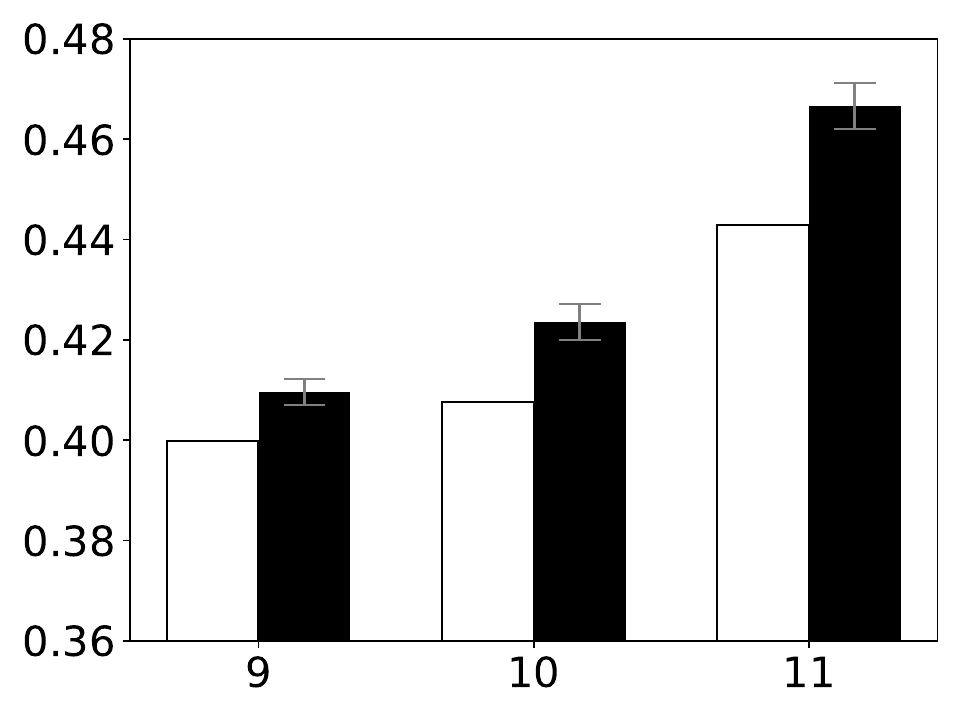}}\,
 \subfigure{\includegraphics[width=0.23\textwidth]{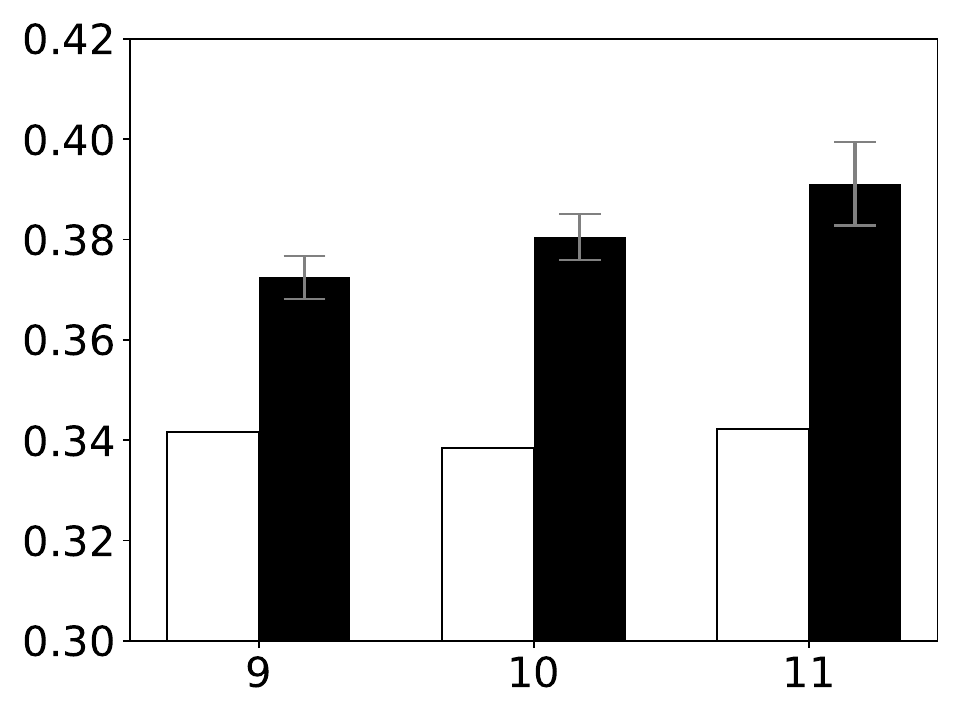}}}
\begin{picture}(0,0)
\put(-40,0){$\scriptstyle N$}
\put(-163,0){$\scriptstyle N$}
\put(-100,75){$\scriptstyle C_\text{mult,norm}$}
\put(-221,75){$\scriptstyle C_\text{mult,norm}$}
\end{picture}
\caption{The multiseed complexity plateau for quantum resonant systems \eqref{QRSdef} with $N=M=9$, 10, 11 and the coupling coefficients \eqref{QRSint} and \eqref{QRScha} for the integrable and chaotic cases, respectively: {\bf (left)} all 0-body operators are included in the seed, {\bf (right)} only the occupation numbers $a^\dagger_ka_k$ are included in the seed. The plotting and normalization conventions are identical to Fig.~\ref{fig:im}. We plot the chaotic value as the average over many realizations ($80, 40, 20$ realizations for $N=M=9$, 10, 11) along with the standard error indicated over the top of the black bars. Smaller values are consistently seen \cite{smallN} for the integrable case (white bars) than for the chaotic case (black bars).\vspace{-3mm}}
\label{fig:qm}
\end{figure}
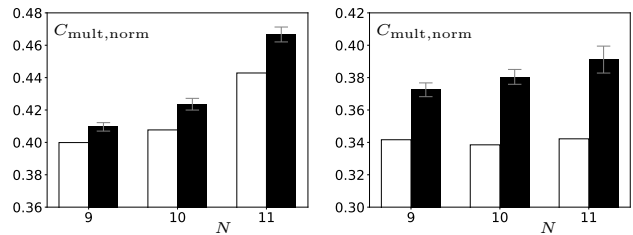
We examine two different sets of interaction coefficients $C$, representing integrable and generic (chaotic) instances of this model:
\beq\label{QRSint}
    C_{nmkl}^{\text{(int)}}=
\begin{cases} 
        0 & \text{if } n \neq 0, m \neq 0, k \neq 0, l \neq 0, \\
        1 & \text{otherwise,}
\end{cases}
\eeq
and
\beq\label{QRScha}
    C_{nmkl}^{(\chi)} \sim U(0,1),
\eeq
where $U(0,1)$ is the uniform random number $\in (0,1)$.
The classical system described by (\ref{QRSint}) is Lax-integrable \cite{trunc}, while for the quantum version, the level spacings of each $(N, M)$-block follow the usual Poisson distribution of integrable systems. For (\ref{QRScha}), the level spacings follow the Wigner-Dyson distribution characteristic of chaotic systems \cite{quantres}. 

It is natural to consider $k$-body operators simple if $k$ is small. The first choice we make for the seed operators are all 0-body operators: those that leave the occupation numbers of individual modes $a_n$ unchanged, as in the Nielsen complexity considerations of  \cite{bound}.
The second option \cite{2body_footnote} is to choose a smaller set consisting only of the number operators for individual modes: $O_k = a_k^\dagger a_k$. 
The performance of the multiseed Krylov complexity plateau (\ref{multiplateau}) for both seed choices is shown in Fig.~\ref{fig:qm}, where lower values are consistently seen for the integrable case (\ref{QRSint}). \vspace{2mm}

To summarize, we have developed an upgrade of Krylov complexity as defined in \cite{K} that takes as its initial seed not a single operator but a collection of all simple operators in the theory \cite{Krav_footnote}, selected according to a straightforward physical criterion (for example, all few-body operators). The late-time plateau of this new quantity, given explicitly by (\ref{multiplateau}), reliably assigns lower values to integrable than to chaotic systems within a set of standard test examples typical of quantum chaos considerations. Our construction can be adapted to deal with state complexity, introduced in \cite{Kstate}, rather than operator complexity, producing similar results, which we briefly review in the supplemental material. There is some apparent similarity between our construction and the notion of `operator size' \cite{size}, where one also starts with a set of simplest operators and grades all other operators in order of increasing complexity. Our precise definition differs, however, and takes as its essential input the actual dynamical evolution via the block Lanczos algorithm in order to develop the operator grading, rather than relying on purely lexicographic criteria applied to the operators written out through the elementary degrees of freedom. We show in the supplemental material that operator size does not reproduce the successful performance of multiseed Krylov complexity demonstrated above in a series of examples.

Besides qualitatively improving the performance of the late-time plateau as an indicator of integrability, our multiseed upgrade of Krylov complexity offers
a conceptual advantage in that the new quantity gives a characterization of the physical theory as such, rather than a characterization of the time evolution of
the chosen seed operator. The way a collection of all simple operators in the theory enters the considerations creates a novel bridge between Krylov and Nielsen complexity as applied to quantum evolution, and more broadly strengthens the contact with computational complexity theory, where the notion of simple elementary operations is of crucial importance.\vspace{2mm}

\begin{acknowledgments} \noindent{\it Acknowledgments.---}
We thank Marine De Clerck and Philip Hacker for discussions and collaboration on related topics, as well as sharing their codes for simulating spin chains \cite{spincode}.
We also thank Adrián Sánchez Garrido for correspondence on numerical implementations of the Lanczos algorithm.
This work has been supported by the Research Foundation Flanders (FWO) through project G012222N, and by Vrije Universiteit Brussel through the Strategic Research Program High-Energy Physics. OE is supported by Thailand NSRF via PMU-B (grant number B13F670063). GP is supported by a PhD fellowship from FWO.
\end{acknowledgments}

\onecolumngrid

\newpage

\begin{centering}
\onecolumngrid
{\large\bf SUPPLEMENTAL MATERIAL\rule{0mm}{7mm}}\vspace{8mm}
\twocolumngrid
\end{centering}

\subsection*{Preferred role of the block Krylov basis}

In \cite{Kstate}, it was shown that the Krylov basis minimizes a cost function quantifying the spread of the initial state over a given basis. We can use similar arguments in the operator (rather than state) complexity context to show that the block Krylov basis minimizes a different but related cost function. In the following, we give the necessary definitions before stating the theorem and giving the proof.

The block Krylov basis $\mathcal{K}_B$ corresponding to an initial seed $\displaystyle \Omega_0 = \big\{ |\mathcal{O}_{0,0}\rangle,|\mathcal{O}_{0,1}\rangle, ...,|\mathcal{O}_{0,m-1}\rangle \big\}$ consists of a sequence of subspaces $\displaystyle \Omega_J$, each having dimension $p_J$. Using these $p_J$, we can take any complete ordered orthogonal basis of operators $\mathcal{B}$ and define a cost function as follows: assign the first $p_0$ vectors a weight of $c_0$, the next $p_1$ a weight of $c_1$ and so on, with $c_J$ positive and increasing. Since this prescription will not necessarily cover a complete basis, we assign the final weight $c_{M-1}$ to every remaining operator in $\mathcal{B}$ instead of just the next $p_{M-1}$. We denote the subspace with weight $c_J$ as $B_J$, and the individual elements of $B_J$ as $|B_{J,k}\rangle$. The cost function is then defined by:
\beq
    \mathcal{C}_\mathcal{B} (t)\! =\! \frac{1}{m} \sum_{J,k,n} c_J |\langle \mathcal{O}_{0,n} (t) | B_{J,k}\rangle |^2 \!= \!\sum_{J} c_J p_\mathcal{B} (J,t),
    \label{costmulti}
\eeq
with
\beq
    p_\mathcal{B} (J,t) = \frac{1}{m} \sum_{k,n} |\langle \mathcal{O}_{0,n} (t) | B_{J,k}\rangle |^2
    \label{prob}
\eeq
representing the average probability, over all elements of $\displaystyle \Omega_0$, to be in subspace $J$ at time $t$. Intuitively, this cost function quantifies the spread of the initial subspace of operators $\displaystyle \Omega_0$ into the sequence of subspaces $B_J$ at time $t$.

Following \cite{Kstate}, we consider the sequence of derivatives of $\mathcal{C}_{\mathcal{B}_1} (t)$ and $\mathcal{C}_{\mathcal{B}_2} (t)$ evaluated at $t=0$, for two different bases $\mathcal{B}_1$, $\mathcal{B}_2$. If the first inequality in this sequence is $\mathcal{C}^{(l)}_{\mathcal{B}_1} (0) < \mathcal{C}^{(l)}_{\mathcal{B}_2} (0)$ then $\mathcal{C}_{\mathcal{B}_1} (t) < \mathcal{C}_{\mathcal{B}_2} (t)$ for any $t$ close enough to $0$. So a functional minimization, valid only at early times, can be defined by comparing the sequence of derivatives at $t=0$.

Equation \eqref{costmulti} is invariant with respect to changes between orthonormal bases if the subspaces $B_J$ remain the same. Given this, consider some complete basis $\mathcal{B}$ for which the $B_J$ are the same as the block Krylov subspaces $\displaystyle \Omega_J$. $\mathcal{B}$ is complete, so the final subspace $B_{M-1}$ will generically contain the corresponding block Krylov subspace $\displaystyle \Omega_{M-1}$, along with some extra elements; the latter are unexplored by the time evolution of $\displaystyle \Omega_0$ since this is completely covered by the block Krylov basis, so they do not affect the final cost. Therefore, we have that $\mathcal{C}_\mathcal{B} (t) = \mathcal{C}_{\mathcal{K}_B} (t)$, allowing us to identify any such basis as equivalent to the block Krylov basis $\mathcal{K}_B$.\vspace{1mm}

\noindent {\bf Theorem:} {\it The block Krylov basis $\mathcal{K}_B$, or any equivalent basis, minimizes the cost function \eqref{costmulti}, with `minimization' understood in the sense described above.}\vspace{1mm}

\noindent {\bf Proof:} The proof follows that of \cite{Kstate}, but we examine the first $N$ subspaces instead of $N$ elements of the basis.
The subspace occupied at $t=0$ is by definition $\displaystyle \Omega_0$. A basis whose first subspace is equal to this, $\displaystyle B_0=\Omega_0$, will have cost $\mathcal{C}^{(0)}_{\mathcal{B}} (0) = c_0$. For any other basis, part of the initial subspace $\displaystyle \Omega_0$ does not appear in $B_0$, so it must appear in some of the later subspaces. This will cause the sum of \eqref{costmulti} to receive contributions from the higher weights, which means the cost for such a basis will be higher at $t=0$.

We can therefore assume the basis $\mathcal{B}$ is such that the first $N>0$ subspaces coincide with those of the block Krylov basis, $\displaystyle B_J = \Omega_J$ for $J=0,1,...,N-1$. If $N=M$, all subspaces coincide and $\mathcal{B}$ has the same cost as $\mathcal{K}_B$. Similarly, if $N=M-1$, the next subspace $B_{M-1}$ will be assigned cost $c_{M-1}$ and will contain the rest of the Hilbert space, including the last remaining block Krylov subspace $\displaystyle \Omega_{M-1}$, so again this will have the same cost as $\mathcal{K}_B$. So we can assume $N<M-1$, and that $\displaystyle B_N \neq \Omega_N$. We can expand the time derivatives of the probabilities appearing in \eqref{costmulti} using $\partial_t |O(t)\rangle = i \mathcal{L}|O(t)\rangle$ as follows:
\begin{align}
    &p^{(l)}_\mathcal{B} (J,t) = \partial_t^l p_\mathcal{B} (J,t) = \frac{i^l}{m}\sum_{k,n} \sum_{j=0}^l (-1)^{l-j}\binom{l}{j}\nonumber\\
     &\hspace{1cm}\times\langle \mathcal{O}_{0,n} (t) |\mathcal{L}^{l-j}| B_{J,k}\rangle \langle B_{J,k} |\mathcal{L}^j| \mathcal{O}_{0,n} (t)\rangle.
    \label{dprob}
\end{align}
For $J<N$, only the subspace $\displaystyle B_J= \Omega_J$ appears in the expression for $p^{(l)}_\mathcal{B} (J,t)$. It follows that $p^{(l)}_\mathcal{B} (J,t) = p^{(l)}_{\mathcal{K}_B} (J,t)$ for all $l$ for $J<N$. Next, we focus on the first $2N-1$ derivatives and look at the contributions from different $J$.

Evaluating \eqref{dprob} at $t=0$, we get terms containing $\mathcal{L}^j| \mathcal{O}_{0,n} (0)\rangle$. For $j<N$, such a term must be entirely contained in the subspaces $\displaystyle B_J= \Omega_J$ with $J<N$, since this is how the block Krylov subspaces are defined. So we must have $\langle B_{J,k} |\mathcal{L}^j| \mathcal{O}_{0,n} (0)\rangle = 0$ for all $J\geq N$ and $j<N$. Because we are considering the first $2N-1$ derivatives, we also have that $l<2N$, so for any integer $j$ we have that either $j<N$ or $l-j<N$. Each term in the sum \eqref{dprob} at $t=0$ is proportional to either $\langle B_{J,k} |\mathcal{L}^j| \mathcal{O}_{0,n} (0)\rangle$ or $\langle B_{J,k} |\mathcal{L}^{l-j}| \mathcal{O}_{0,n} (0)\rangle$, so it vanishes if $J\geq N$ and $l<2N$. Combining the results we have so far, $p^{(l)}_\mathcal{B} (J,t) = p^{(l)}_{\mathcal{K}_B} (J,t)$ for all $J$ and any $l<2N$. To see which basis minimizes the cost function at early times, we therefore have to examine the next derivative.

Taking $l=2N$, we can again use the fact that $\displaystyle B_J= \Omega_J$ for $J<N$ to eliminate most of the terms in \eqref{dprob} at $t=0$. The only remaining term corresponds to $j=l-j=N$:
\begin{align}
    &p^{(2N)}_\mathcal{B} (J,0) =\frac{1}{m} \sum_{k,n}  \binom{2N}{N} \nonumber \\
    &\hspace{1cm}\times\langle \mathcal{O}_{0,n} (0) |\mathcal{L}^N| B_{J,k}\rangle \langle B_{J,k} |\mathcal{L}^N| \mathcal{O}_{0,n} (0)\rangle.
\end{align}
There can be a component of $\mathcal{L}^N| \mathcal{O}_{0,n} (0)\rangle$ which is orthogonal to all the subspaces $B_0, ..., B_{N-1}$, call it $|X_n\rangle$. Because $\displaystyle B_N \neq \Omega_N$, we must have $\langle X_n | X_n \rangle > 0$ for some $n$. Since the basis $\mathcal{B}$ is orthogonal, the only contributions to the probability that can appear for $J\geq N$ are from the $|X_n \rangle$:
\beq
    p^{(2N)}_\mathcal{B} (J\geq N,0) = \frac{1}{m} \sum_{k,n} \binom{2N}{N} \langle X_n| B_{J,k}\rangle \langle B_{J,k} |X_n\rangle.
\eeq
Given that for $J<N$ the probabilities are equal to those of the block Krylov basis, we have every term we need in order to compare the $2N$ derivatives of the cost function:
\begin{align}
    \mathcal{C}&_\mathcal{B}^{(2N)} (0) = \sum_{J} c_J p_\mathcal{B}^{(2N)} (J,0) = \sum_{J=0}^{N-1} c_J p_\mathcal{B}^{(2N)} (J,0) \nonumber \\
    &+ \frac{1}{m} \sum_{k,n}  \sum_{J=N}^{M-1}\binom{2N}{N} c_J \langle X_n| B_{J,k}\rangle \langle B_{J,k} |X_n\rangle.
    \label{2nder}
\end{align}
In the first sum, we can replace all derivatives of the probabilities with the corresponding block Krylov ones, since $J<N$. Furthermore, in the second term, we can replace all the $c_J$ with just $c_N$ to get the following inequality:
\begin{align}
     \mathcal{C}_\mathcal{B}^{(2N)} (0) &\geq \sum_{J=0}^{N-1} c_J p_{\mathcal{K}_B}^{(2N)} (J,0)  \nonumber \\
     &+\frac{c_N}{m} \binom{2N}{N}\sum_{k,n}  \sum_{J=N}^{M-1} \langle X_n| B_{J,k}\rangle \langle B_{J,k} |X_n\rangle.
     \label{ineqder}
\end{align}
The only way to get equality is when only $c_N$ contributes to the second sum of \eqref{2nder}. This happens only if all the $|X_n\rangle$ are contained in the subspace $B_N$, which is not the case because we are assuming $\displaystyle B_N \neq \Omega_N$. We therefore have a strict inequality. We have already argued that the $|X_n\rangle$ only have support in $B_J$ for $J\geq N$, so in the last term of \eqref{ineqder}, the sums over $J,k$ represent the full inner product of each $|X_n\rangle$:
\beq
     \mathcal{C}_\mathcal{B}^{(2N)} (0) > \sum_{J=0}^{N-1} c_J p_{\mathcal{K}_B}^{(2N)} (J,0) + \frac{c_N}{m} \binom{2N}{N}\sum_n  \langle X_n |X_n\rangle.
\eeq
These are also the terms that appear in $\mathcal{C}_{\mathcal{K}_B}^{(2N)} (0)$, since all the $|X_n\rangle$ are by definition contained in $\displaystyle \Omega_N$, and will be assigned weight $c_N$. So we have:
\beq
    \mathcal{C}_\mathcal{B}^{(2N)} (0) > \mathcal{C}_{\mathcal{K}_B}^{(2N)} (0),
\eeq
which means the cost function for $\mathcal{B}$ is greater than that for $\mathcal{K}_B$ at early times.

\subsection*{Multiseed complexity of spread of states}

Spread complexity, first proposed in \cite{Kstate},  provides a modification of the Krylov complexity of \cite{K}, formulated in terms of the evolution of quantum states rather than operators. The idea is to start with a generic measure of the spread of the wavefunction over the Hilbert space relative to some arbitrary basis $|B_j\rangle$:
\beq
    C_{B} = \sum_{j=0}^{K-1} c_j |\langle \psi_0 (t)|B_j\rangle |^2,
\label{costf}
\eeq
with $c_j$ positive and increasing. It is then argued that appropriately minimizing this cost function over all possible bases uniquely yields the Krylov basis.

The Lanczos algorithm can be applied to quantum states, with the Hamiltonian substituted for the Liouvillian. The coefficients $a_j$ in the analog of (1) in the main text no longer necessarily vanish. This can again be used to iteratively construct the Krylov basis $|\psi_j\rangle$ starting from an initial state $|\psi_0\rangle$. Spread complexity is then given by:
\beq
    C_S (t) = \sum_{j=0}^{K-1} j |\langle \psi_0 (t)|\psi_j\rangle |^2,
\eeq
obtained from \eqref{costf} in the Krylov basis by setting the cost sequence to $c_j=j$, so that spread complexity represents the average `position' of the wavefunction in terms of the sequential numbering of the Krylov basis vectors. This quantity behaves similarly to Krylov complexity, displaying initial growth and eventually plateauing at late times.

For the multiseed variant, if the seed is some collection of states $\displaystyle {\Psi}_0$, the block Krylov basis can be constructed by orthogonalizing $\displaystyle \big\{{\Psi}_0, {H}{\Psi}_0, {H^2}{\Psi}_0, ...\big\}$. The block Lanczos algorithm can still be used to streamline the construction. This basis allows us to again cover the time evolution of each seed state $|\psi_{0,n}\rangle$:
\beq
    |\psi_{0,n}(t)\rangle = e^{-iHt} |\psi_{0,n}\rangle = \sum_{J=0}^{M-1} \sum_{k=0}^{p_J-1} \phi_{J,k}^{(n)} (t) |\psi_{J,k} \rangle,
\label{tkrylst}
\eeq
which is the analog for quantum states of the operator-based considerations in the main text. The {\it multiseed state complexity} is defined as the quantum-mechanical average of the basis level number for a given initial wavefunction, further averaged over all the level 0 seeds used to initialize the block Lanczos process. Its all-time average is once again easily obtained by introducing the eigenbasis of $H$, denoted $|E_\alpha\rangle$ with corresponding eigenvalues $E_\alpha$:
\begin{align}
\overline{C_{\text{mult,state}}} &= \frac{1}{m}\sum_{n=0}^{m-1} \sum_{J=0}^{M-1} \sum_{k=0}^{p_J-1}\!\!\sum_{\substack{\alpha,\beta \\ E_\alpha = E_\beta}}\!\! J \langle \psi_{0,n} | E_\alpha \rangle \langle E_\beta | \psi_{0,n} \rangle \nonumber\\
    &\hspace{1.5cm}\times \langle \psi_{J,k} | E_\beta \rangle \langle E_\alpha | \psi_{J,k} \rangle.
\label{multiplateaust}
\end{align}

\begin{figure}[t]
\centering
\mbox{\subfigure{\includegraphics[width=0.23\textwidth]{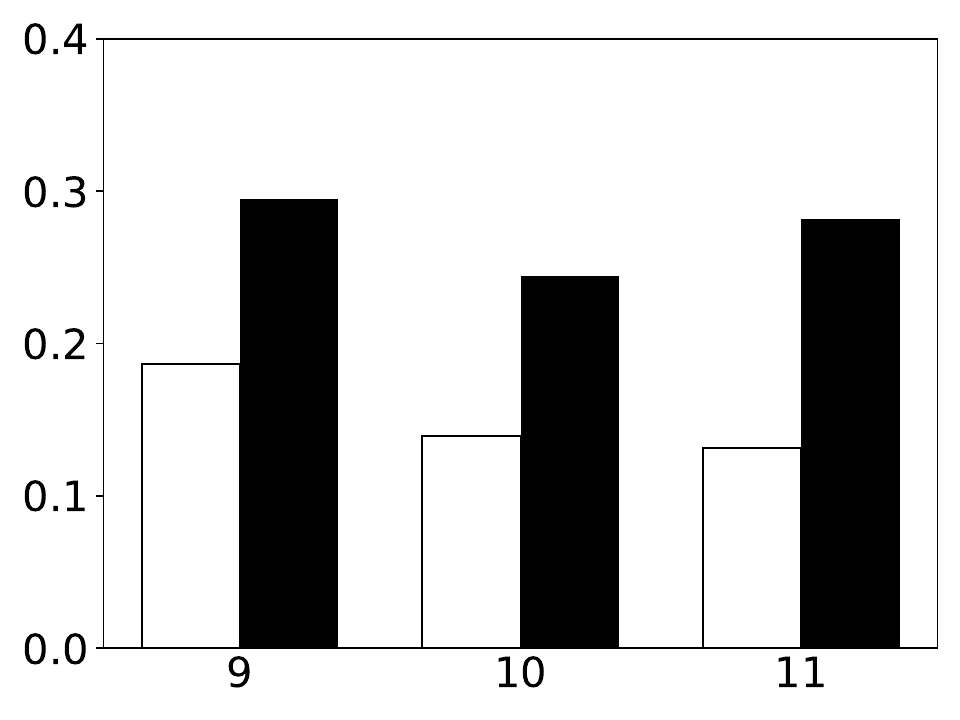}}\,
 \subfigure{\includegraphics[width=0.23\textwidth]{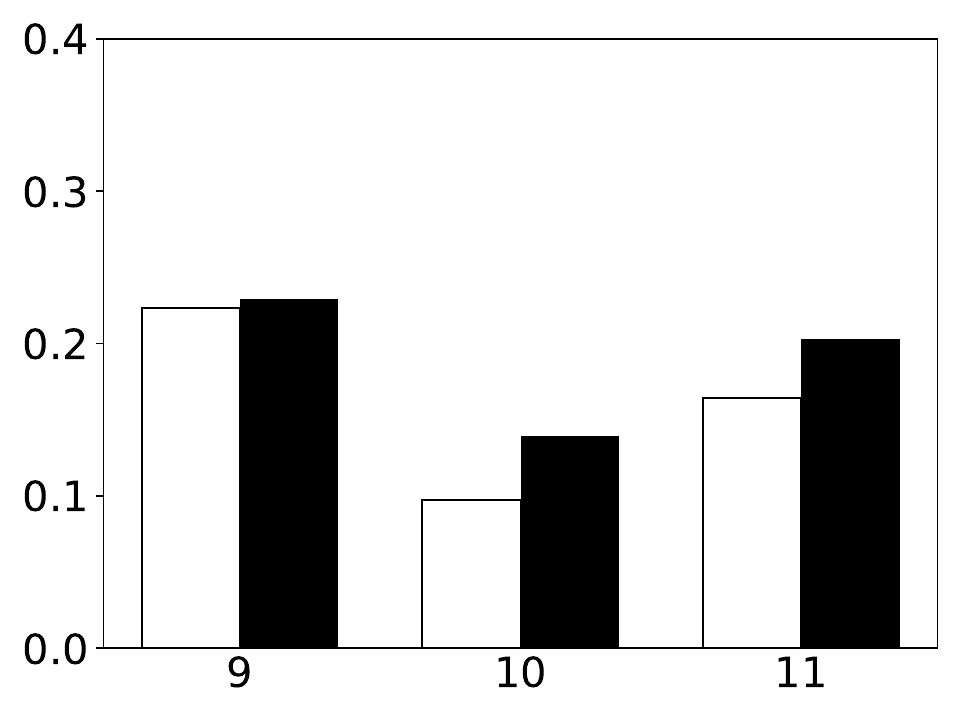}}}
\begin{picture}(0,0)
\put(-40,0){$\scriptstyle L$}
\put(-163,0){$\scriptstyle L$}
\put(-104,75){$\scriptstyle C_\text{mult,norm}$}
\put(-224,75){$\scriptstyle C_\text{mult,norm}$}
\end{picture}
\caption{Multiseed state complexity plateau {\bf(left)} for the Ising chain and {\bf(right)} for the XYZ chain. The plotting conventions are identical to the main text.
Smaller values are seen for the integrable cases (white bars) than for the chaotic cases (black bars).\vspace{-3mm}}
\label{fig:ims}
\end{figure}
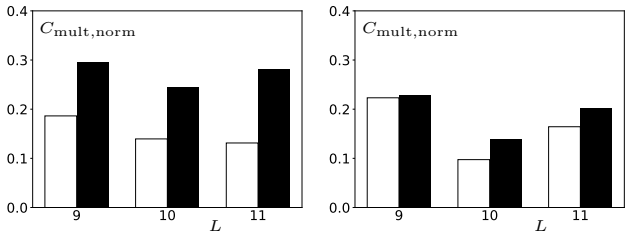
One complication relative to the operator setting is that it is much less straightforward to decide which states are simple on the basis of some physical reasoning (there is no immediate analog of few-body operators). We have nonetheless observed that the formalism works well for spin chains with a natural, even if somewhat {\it ad hoc}, choice of the simple states. Namely, we consider product states where each particle has a definite spin in the $x$, $y$ or $z$ direction. Of these, we select the following subset: the six states where all spins are the same, and all states where all the spins point in the same direction except for one, pointing in the opposite direction. 
We show the performance of \eqref{multiplateaust} with these specifications in Fig.~\ref{fig:ims}. This formulation reliably assigns smaller values to the integrable cases.

\subsection*{Operator size}

Another quantity that has appeared in operator complexity considerations for quantum systems is the operator size \cite{size}, which measures how many `simple' operators are involved in the mathematical expression for the chosen operator. Here, `simple' means acting on only one particle or site. More concretely, one starts with a basis of operators where an element $B_{J,k}$ is expressed as a product of $J$ simple operators. This prescription splits the basis into different sets (indexed by $J$), where every member (indexed by $k$) of each set has the same cost. Using this, the operator size of $\mathcal{O}$ is given by
\beq
    s(\mathcal{O}) = \sum_{J, k} J \big|\Tr[ \mathcal{O}^\dagger B_{J,k}]\big|^2\equiv \sum_{J, k} J |\langle \mathcal{O}|B_{J,k}\rangle |^2.
\eeq

The operator size is connected to Krylov complexity via the framework of {\it q-complexities} introduced in \cite{K}. Other than these two examples, this class of quantities also includes other measures of operator complexity under current investigation, including out-of-time-order correlators (OTOCs). Within this paradigm, Krylov complexity is especially important, as it was shown that it provides an upper bound on the growth of any q-complexity. (Note that this, in particular, implies that standard Krylov complexity would provide an upper bound on the multiseed complexity evaluated for individual seed operators. This upper bound does not, however, share the successful performance of the full multiseed complexity for distinguishing integrability and chaos, as evident from our exposition in the main text.)

The difference between this construction and our multiseed complexity is that the operator size receives no input from the detailed specification of the Hamiltonian, and in this sense is purely kinematic rather than dynamical. The entire basis from simple to complex operators has to be specified manually, based for example on the locality or rank of the operators. Since our goal is to characterize the general properties of a given Hamiltonian, we collect the simplest operators of the system and consider their late-time sizes, similarly to what was done in the main text for multiseed Krylov complexity. We start by considering the $m$ simplest operators with $J=1$, and compute each of their sizes under time evolution:
\beq
    s(B_{1,l}(t)) = \sum_{J, k} J |\langle B_{1,l}|e^{-i\mathbf{\mathcal{L}}t}|B_{J,k}\rangle |^2.
\label{sizesimple}
\eeq
We collect these in a single quantity
\beq
    s_\text{simple} (t) = \frac{1}{m}\sum_{l=1}^{m} s(B_{1,l}(t)),
\label{sizeavg}
\eeq
and finally calculate its all-time average, which again effectively matches the late-time plateau:
\begin{align}
\overline{s_\text{simple}} &= \frac{1}{m}\sum_{l=1}^{m} \sum_{J,k}\!\!\sum_{\substack{\alpha,\beta \\ \omega_\alpha = \omega_\beta}}\!\! J \langle B_{1,l} | \omega_\alpha \rangle \langle \omega_\beta | B_{1,l} \rangle \nonumber\\
    &\hspace{1.5cm}\times \langle B_{J,k} | \omega_\beta \rangle \langle \omega_\alpha | B_{J,k} \rangle.
\label{sizeplateau}
\end{align}
This directly parallels our construction of multiseed Krylov complexity, but with a different grading on the space of operators.
Again, we try to cancel the scaling of this quantity with system size by dividing the plateau by the highest weight assigned to the operators, resulting in a quantity that is always between 0 and 1.

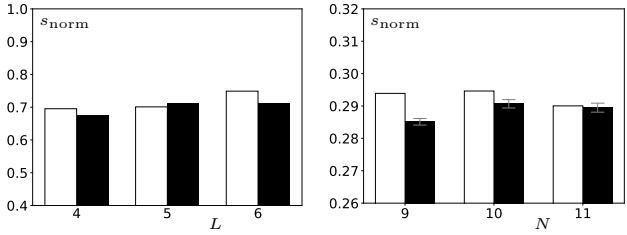
\begin{figure}[t]
\centering
\mbox{\subfigure{\includegraphics[width=0.23\textwidth]{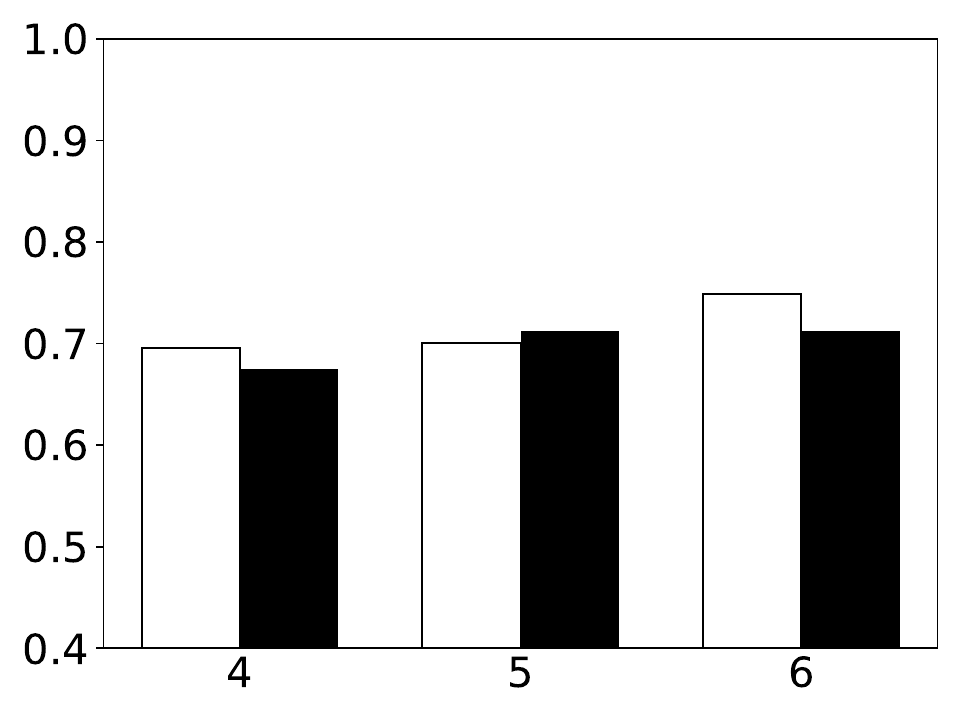}}\,
 \subfigure{\includegraphics[width=0.23\textwidth]{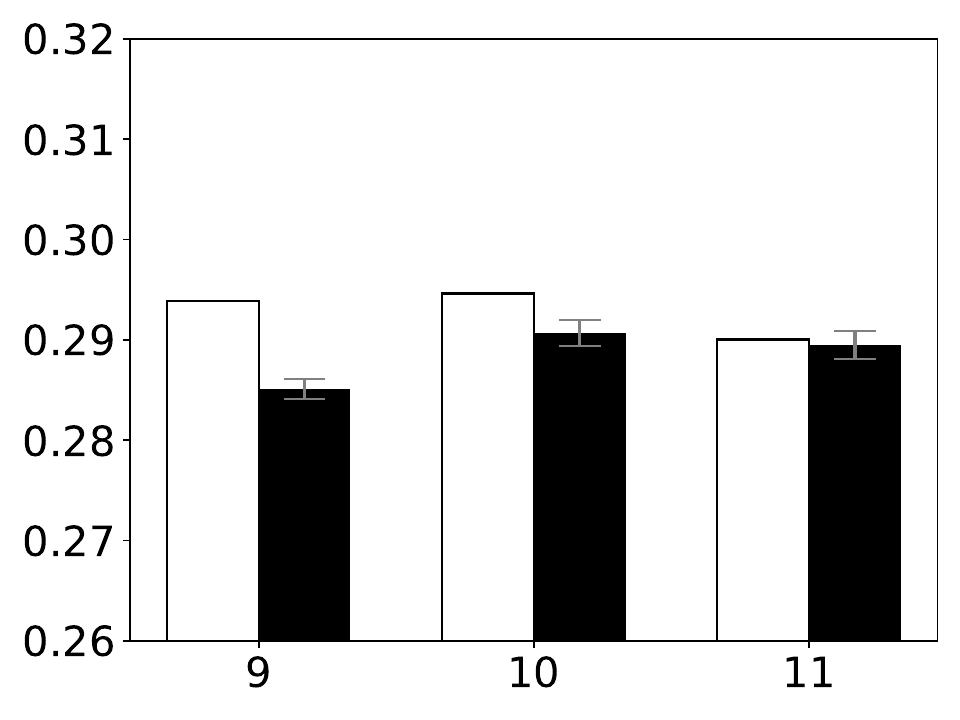}}}\vspace{-3mm}
\begin{picture}(0,0)
\put(-40,0){$\scriptstyle N$}
\put(-163,0){$\scriptstyle L$}
\put(-101,77){$\scriptstyle s_\text{norm}$}
\put(-224,77){$\scriptstyle s_\text{norm}$}
\end{picture}
\caption{The normalized operator size plateau for {\bf (left)} the XYZ chain, {\bf (right)} quantum resonant systems. The integrable cases are not correctly identified in the sense that higher complexity values (white bars) are assigned to them than to the chaotic cases (black bars). This shows the difference in the performance of operator size and multiseed Krylov complexity.\vspace{-3mm}}
\label{fig:xos}
\end{figure}
For spin chains, a basis of operators is constructed from
\beq
S_{a_1}^{(j_1)} S_{a_2}^{(j_2)} ...S_{a_n}^{(j_n)},
\eeq
where $S_{a}^{(j)}$ is the $a^{\text{th}}$ Pauli matrix acting on site $j$ and $1\leq j_1<j_2<...<j_n\leq L$, with $L$ being the number of sites as above and $0\leq n \leq L$. It is natural to distinguish the members of this basis in terms of $n$, with more complex operators having larger $n$.
For quantum resonant systems, a state can be described in terms of the usual Fock basis, so we can form an operator basis by listing all the operators that connect two different Fock states within a given $(N, M)$-block. We then group them based on their rank, which is the number of single-unit changes they make in  the occupation number, with simpler operators changing the occupation numbers less than more complex ones.

We have observed that this construction, which is the operator size analog of what we did for the multiseed Krylov complexity in the main text, works correctly for Ising chains by assigning smaller complexity values to the integrable case (though the difference between integrable and chaotic cases is very small). By contrast, for the XYZ chain and quantum resonant systems, the operator size fails and assigns higher complexity to the integrable cases, as seen in Fig.~\ref{fig:xos}. This shows that, despite some similarities, the successful performance of multiseed Krylov complexity as an integrability measure is not shared by the operator size.

\subsection*{Averaged single-seed complexity}

\begin{figure}[b]
\centering
\mbox{\subfigure{\includegraphics[width=0.23\textwidth]{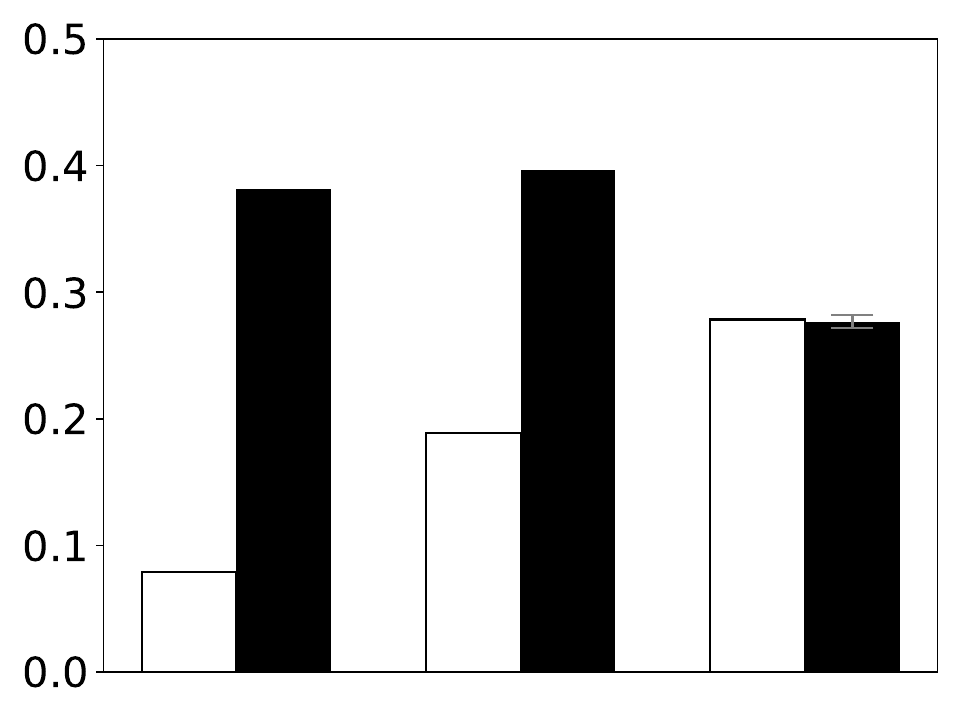}}}
\begin{picture}(0,0)
\put(-30,-5){QRS}
\put(-67,-5){XYZ}
\put(-100,-5){Ising}
\put(-104,73){averaged $\overline{C_{K}}$}
\end{picture}
\caption{Normalized ordinary Krylov complexity plateau averaged over 1-body operators for Ising and XYZ chains, both with $L=4$, and for quantum resonant systems (QRS) with $N=M=9$. For each integrable-chaotic (white-black) pairing, we divide both results by the highest iterative level reached by the Lanczos algorithm (maximized over all seeds, trials, and jointly maximized over the integrable and chaotic cases), which puts all the heights into the interval $(0,1)$.}
\label{fig:Kav}
\end{figure}

In the definition of multiseed complexity, we developed a grading of operators based on the block Lanczos algorithm (qualitatively, in order of increasing complexity), then introduced the quantum-mechanical average of the position with respect to this graded basis for trajectories starting with individual operators at the lowest (simplest) level, and then averaged over all simple initial operators.
It is a natural thought that one could apply the same construct to the original Krylov complexity of \cite{K}, computing it for each simple operator and then averaging over all such operators.

Before exploring this quantity any further, we remark that it suffers from a number of considerable disadvantages. First, it is much more numerically demanding that our multiseed construction. In the multiseed construction, one runs the (block) Lanczos algorithm only once. For the averaged Krylov complexity, one will have to rerun it for each operator in the simple set, which is a huge number of runs. For instance, the number of 1-body operators is, roughly, the number of degrees of freedom, which by definition diverges in the thermodynamic limit. 

Even more dramatically, the Krylov complexity average is sensitive to the choice of basis in the space of simple operators, which is not the case for multiseed complexity. The latter comes from two key properties: first, the subspaces generated by the block Lanczos algorithm depend only on the span of the initial seed, not on the specific basis; second, the multiseed complexity also only depends on these subspaces. Specifically, the expression involves a sum of projections over some basis within each subspace, ensuring that the result is basis-independent. This appealing property of multiseed complexity (which resolves some of the ambiguity inherent in standard Krylov complexity) is not shared by the average of the single-seed Krylov complexity (which would require an explicit basis specification as its input). For example, we found that simply rotating the $(x,y,z)$ axes for the seed operators in a spin chain, while not affecting the multiseed complexity, can drastically alter the averaged Krylov complexity.

Despite these flaws of the averaged-Krylov construction, we find it useful to report the results here, for the relatively small systems we can access numerically: the averaged Krylov complexity plateau distinguishes integrable and chaotic cases correctly for spin chains, but not for quantum resonant systems, as seen in Fig.~\ref{fig:Kav}. (As our simple operator basis we took the explicit sets of operators described in our considerations of multiseed complexity, with the same spatial axes used in the expressions for the Hamiltonians.)


\begin{thebibliography}{99}
\bibitem{Frob}Other choices are possible, but this Frobenius inner product is simple and convenient, and has been common in Krylov complexity studies starting with \cite{K}.
\bibitem{smallN}We have observed that the integrable-chaotic hierarchy is actually reversed for the smaller blocks with $N=M=7$ and 8. However,
the dimensions of these blocks are 15 and 22, respectively, and no quantum chaos measure is expected to perform reliably for such tiny Hilbert spaces.
\bibitem{2body_footnote}It would also be interesting to consider more complex seeds, say, all 2-body operators. One is however strongly limited by the numerical costs. For Hilbert space sizes we can access, the 2-body operators take up of order 10\% of all operators. There are hence only a few block Lanczos algorithm levels, not enough for the construction to work well. (Nielsen complexity performance has been explored in depth in relation to different operator complexity thresholds in \cite{bound,complint}.) If we ignore this issue and proceed, the result is that the 2-body multiseed complexity correctly distinguishes integrability and chaos for spin chains, but not for quantum resonant systems. We expect that the latter will be rectified for bigger $(N,M)$ blocks than what we can currently access, but the matter will require further investigation.
\end{thebibliography}
\end{document}